\documentclass[12pt,a4paper]{article}
\usepackage{t1enc}
\usepackage[latin1]{inputenc}
\usepackage[english]{babel}
\usepackage{graphics}

\begin{document}

\addtocounter{footnote}{1}

\title{Example of a self-consistent solution for a fermion on domain wall}

\addtocounter{footnote}{0}

\author{V. A. Gani$^{1,2}$, V. G. Ksenzov$^2$, A. E. Kudryavtsev$^2$\\
$^{1}$\it\small Department of Mathematics, National Research Nuclear University ''MEPhI'',\\
\it\small Moscow 115409, Russia.\\
$^{2}$\it\small State Research Center Institute of Theoretical and Experimental Physics,\\
\it\small Moscow 117218, Russia.\\
}

\date{}

\maketitle

\begin{abstract}

We discuss a self-consistent solution for a fermion coupled to static scalar field 
in the form of a kink (domain wall). In particular, we study the case when the fermion
occupies an excited non-zero frequency level in the presence of the domain wall field.
The effect of the domain wall profile distortion is calculated analytically.

\end{abstract}

\section{Introduction}

The problem of the spectrum of a fermion coupled to the field of a static kink was discussed for the first time in the classical paper by Dashen, Hasslacher and Neveu~\cite{DHN}.
Some details of the problem were later discussed in Refs.~\cite{JR} and \cite{GW}. The fermion spectrum and scattering states in the presence of a domain wall were
studied in Refs.~\cite{Vol} and \cite{Stojk}. The details of the problem were reviewed in Radjaraman's book~\cite{Rad}.

In particular, these papers were devoted to the problem of a zero-frequency fermion coupled to the kink.
As far as we know, no other exact analytical solution for the problem ''domain wall + fermion'' is known at the present time.
The reason for this is that when dealing with non-zero fermionic excitations one needs to account for the distortion of the profile of kink in response to the coupling of fermion.
Note that this distortion depends on coupling constant and hence is not small in general, implying that the problem ''fermion in the field of a kink'' is to be solved in a self-consistent way, taking
into account the kink's distortion due to the presence of coupled fermion.
Below we do not consider the effect of ''fermion loop''.
This is in line with the arguments given in Ref.~\cite{DHN}. Actually, the authors of Ref.~\cite{DHN} look for time-independent (static) solutions
for scalar field $\phi$. Variating the effective action for the time-independent but spatially varying field $\phi$,
they managed to obtain an effective equation of motion for the field $\phi$ with separate contribution from ''occupied states'' and ''fermion loop''
(see also eq. (4.10) of paper~\cite{DHN} and the discussion related). Following the arguments of Ref.~\cite{DHN}, we shall concentrate on the effect of ''occupied states'' only,
i.e. shall work in Hartree-Fock type approximation.
The effect of ''fermion loop'' is a quantum mechanical correction of order $\hbar$, and we shall ignore the effect of ''fermion loop'' in this paper.

Indeed, the self-consistent treatment of the problem ''domain wall + excited fermion''
is rather complicated even at the Hartree-Fock level. However, as we shall demonstrate, for some special values of meson-fermion coupling the problem of an excited fermionic state
in the field of a kink may be solved in a self-consistent analytical form.

Note that a similar problem, ''kink + charged scalar field'' was solved in our paper~\cite{Lensky}.
The related problem of the interaction of a domain wall with a skyrmion was studied in Ref.~\cite{KPZ} .

The plan of our paper is the following. In {\bf Section 2} we discuss the equations of motion and give analytic solutions for both ground state (zero
mode) and first excited modes of the fermion in the field of domain wall. In {\bf Section 3} a self-consistency equation is formulated and solved.
Finally, a general discussion and a summary of the results are presented in {\bf Section 4}.

\section{Lagrangian, equation of motion and some solutions}
\setcounter{equation}{0}

We study the system of the interacting scalar ($\phi$) and fermion ($\Psi$) fields in two-dimensional space-time (1+1).
The corresponding Lagrangian density is
\begin{equation}
\mathcal{L}=\frac{1}{2}\left(\partial_\mu\phi\right)^2-\frac{m^2}{2}\left(\phi^2-1\right)^2
+\bar{\Psi}i\hat{\partial}\Psi-g\bar{\Psi}\Psi\phi.
\label{lagrangian}
\end{equation}
Our notation here is similar but not identical to that of Refs.~\cite{DHN,JR,Rad}.
Scalar field $\phi(x,t)$ is dimensionless, $[\phi]=1$. The dimension of fermionic field $\Psi$ is $[\Psi]=m^{1/2}$, where $m$ is mass parameter.
Coupling constant $g$ is also dimensionful, $[g]=m$.

Introducing space-time dimensionless variables $\tilde x=mx$, $\tilde t=mt$ and rescaling fermionic field $\tilde\Psi=m^{-1/2}\Psi$ and the coupling constant $\tilde g=m^{-1}g$,
we rewrite the Lagrangian density (\ref{lagrangian}) in the form ${\cal L}(x,t)=m^2\tilde{\cal L}(\tilde x,\tilde t)$, where
\begin{equation}
\tilde{\cal L}=\frac{1}{2}\left(\tilde\partial_\mu\phi\right)^2-\frac{1}{2}\left(\phi^2-1\right)^2
+\tilde{\bar{\Psi}}i\tilde{\hat{\partial}}\tilde{\Psi}-\tilde{g}\tilde{\bar{\Psi}}\tilde{\Psi}\phi.
\label{lagr_new}
\end{equation}
Hereinafter we omit everywhere the tilde and work in terms of dimensionless fields, coupling constant and space-time variables.

If the coupling of scalar field to fermions is switched off, $g=0$, the equation of motion for scalar field reads:
\begin{equation}
\partial_\mu\partial^\mu\phi-2\phi+2\phi^3=0.
\label{eqmo1}
\end{equation}
The constant solutions of (\ref{eqmo1}) $\phi_{\pm}=\pm 1$ correspond to the degenerate absolute minima of the Hamiltonian $H[\phi]$ (vacuum states) and the solution
$\phi=0$ corresponds to the unstable vacuum, a state with non-violated symmetry.
Another static solution of (\ref{eqmo1}) with finite energy is a topological solitary wave called ''kink'', 
\begin{equation}
\phi_K(x,t)=\tanh x.
\end{equation}
In three space dimensions, this solution looks like a domain wall that separates two space regions with different vacua $\phi_{\pm}$. The kink is an extended
object with a ground state and a set of excited states above it, see also Refs.~\cite{BelK,Rub}.

Let us discuss the fermionic sector of the theory. After the substitution $\Psi(x,t)=e^{-i\varepsilon t}\psi_\varepsilon(x)$ the Dirac equation for the massless
case reads (see, e.g.~\cite{Rad}):
\begin{equation}
\left(\varepsilon+i\alpha_x\frac{\partial}{\partial x}-g\beta\phi(x)\right)\psi_\varepsilon(x)=0,
\label{eq_psi}
\end{equation}
where $\alpha_x$ and $\beta$ are the Pauli matrices,
\begin{equation}
\alpha_x=
\left(
\begin{array}{lr}
0 & -i\\
i & 0
\end{array}
\right),\quad
\beta=
\left(
\begin{array}{lr}
0 & 1\\
1 & 0
\end{array}
\right).
\end{equation}

In eq.~(\ref{eq_psi}),
\begin{equation}
\psi_\varepsilon(x)=
\left(
\begin{array}{l}
u_\varepsilon(x)\\
v_\varepsilon(x)
\end{array}
\right)
\end{equation}
is the two-component wave function of fermion. In terms of functions $u_\varepsilon(x)$ and $v_\varepsilon(x)$, eq.~(\ref{eq_psi}) takes the form
\begin{equation}
\left\{
\begin{array}{l}
\displaystyle\frac{du_\varepsilon}{dx}+g\phi(x)u_\varepsilon=\varepsilon v_\varepsilon,\\
\\
-\displaystyle\frac{dv_\varepsilon}{dx}+g\phi(x)v_\varepsilon=\varepsilon u_\varepsilon.
\end{array}
\right.
\label{system1}
\end{equation}
This system of equations may be presented in the form
\begin{equation}
\left\{
\begin{array}{l}
-\displaystyle\frac{d^2u_\varepsilon}{dx^2}+\left(g^2\phi^2(x)-g\frac{d\phi}{dx}\right)u_\varepsilon=\varepsilon^2 u_\varepsilon,\\
\\
-\displaystyle\frac{d^2v_\varepsilon}{dx^2}+\left(g^2\phi^2(x)+g\frac{d\phi}{dx}\right)v_\varepsilon=\varepsilon^2 v_\varepsilon.
\end{array}
\right.
\label{system2}
\end{equation}
Substituting $\phi(x)=\phi_K(x)=\tanh x$ we finally get:
\begin{equation}
\left\{
\begin{array}{l}
-\displaystyle\frac{d^2u_\varepsilon}{dx^2}-\frac{g(g+1)}{\cosh^2 x}u_\varepsilon=(\varepsilon^2-g^2) u_\varepsilon,\\
\\
-\displaystyle\frac{d^2v_\varepsilon}{dx^2}-\frac{g(g-1)}{\cosh^2 x}v_\varepsilon=(\varepsilon^2-g^2) v_\varepsilon.
\end{array}
\right.
\label{system3}
\end{equation}

This system describes the spectrum and eigenfunctions of the fermion in the external scalar field of kink $\phi_K(x)=\tanh x$.
Let us look at some simple analytic examples of solutions for system~(\ref{system3}).

{\bf 1.} \underline{\bf The case $g=1$}. The spectrum in this case consists of only one localized nondegenerate bound state with $\varepsilon=0$
(zero frequency mode) plus the continuum of scattering states with $\varepsilon^2\ge 1$. The normalized fermion bound state wave function is
\begin{equation}
\Psi^{g=1}_{\varepsilon=0}(x,t)=\frac{1}{\sqrt{2}}
\left(
\begin{array}{c}
\displaystyle\frac{1}{\cosh x}\\
\\
0
\end{array}
\right).
\label{psi_g1_e0}
\end{equation}

{\bf 2.} \underline{\bf The case $g=2$}. There are two different bound states in this case: one with $\varepsilon=0$ and one with $\varepsilon^2=3$.
The continuum of scattering states corresponds to $\varepsilon^2\ge 4$.

{\bf 2A.} The ground state of the system in this case is a nondegenerate state with eigenvalue $\varepsilon=0$. The wave function in this case is
\begin{equation}
\Psi^{g=2}_{\varepsilon=0}(x,t)=\frac{\sqrt{3}}{2}
\left(
\begin{array}{c}
\displaystyle\frac{1}{\cosh^2 x}\\
\\
0
\end{array}
\right).
\label{psi_g2_e0}
\end{equation}

{\bf 2B.} Nonzero frequencies for bound states appear in couples and hence they are twicely degenerate. In our case they correspond to $\varepsilon=\pm\sqrt{3}$.
The states with positive frequencies are fermions and with negative ones are antifermions.
The wave functions of these states are
\begin{equation}
\Psi^{g=2}_{\varepsilon=\pm\sqrt{3}}(x,t)=e^{\mp i\sqrt{3}t}
\left(
\begin{array}{c}
\pm\displaystyle\frac{\sqrt{3}}{2}\frac{\tanh x}{\cosh x}\\
\\
\displaystyle\frac{1}{2}\:\frac{1}{\cosh x}
\end{array}
\right).
\label{psi_g2_e3}
\end{equation}

Note that negative coupling $g<0$ results in the same spectrum of the problem as positive $g>0$.
The wave functions are similar to those given by (\ref{psi_g1_e0})-(\ref{psi_g2_e3}), with the exchange $u_\varepsilon\leftrightarrow v_\varepsilon$.

\section{One example of nontrivial self-consistent solution}

\setcounter{equation}{0}

The equation of motion for scalar field $\phi(x,t)$ in presence of fermionic field $\Psi$ reads
\begin{equation}
\partial_\mu\partial^\mu\phi-2\phi+2\phi^3=-g\bar{\Psi}\Psi,
\label{eqmo2}
\end{equation}
where $\bar{\Psi}=\Psi^{\dag}\beta$.

Zero frequency solutions (\ref{psi_g1_e0}) and (\ref{psi_g2_e0}) satisfy automatically the condition
$\bar{\Psi}\Psi=2u_{\varepsilon=0}v_{\varepsilon=0}\equiv 0$. Thus, the solution in the form ''$\phi_K(x)=\tanh x$ plus zero mode bound fermion'' is self-consistent.
However, if the fermion occupies a level with $\varepsilon\neq 0$, the r.h.s. of eq.~(\ref{eqmo2}) is already nonzero.
Hence the kink's profile has to be modified to fulfill eq. (\ref{eqmo2}).

As we shall demonstrate below, for some exceptional values of coupling constant $g$ the kink's profile consistent with the fermion field coupled to the kink can be found analytically.

In fact, let us look for a solution of the Dirac equation for the fermion in the field of distorted kink $\tilde{\phi}_K=\tanh{\alpha x}$, where $\alpha$ is unknown real
parameter to be determined from a self-consistency condition that will be derived below. Introducing new variable $y=\alpha x$ and a new parameter $s=g/\alpha$, we get a system of equations
for the fermionic wave function, which formally coincides with (\ref{system3}):
\begin{equation}
\left\{
\begin{array}{l}
-\displaystyle\frac{d^2u_{\varepsilon^\prime}}{dy^2}-\frac{s(s+1)}{\cosh^2 y}u_{\varepsilon^\prime}=({\varepsilon^\prime}^2-s^2) u_{\varepsilon^\prime},\\
\\
-\displaystyle\frac{d^2v_{\varepsilon^\prime}}{dy^2}-\frac{s(s-1)}{\cosh^2 y}v_{\varepsilon^\prime}=({\varepsilon^\prime}^2-s^2) v_{\varepsilon^\prime},
\end{array}
\right.
\label{system4}
\end{equation}
where $\varepsilon^\prime=\varepsilon/\alpha$. The solution for the case $\varepsilon^\prime=\sqrt{3}$, $s=2$ is
\begin{equation}
\Psi^{s=2}_{\varepsilon^\prime=\sqrt{3}}(x,t)=e^{-i\sqrt{3}\alpha t}
\left(
\begin{array}{c}
\displaystyle\frac{\sqrt{3\alpha}}{2}\frac{\tanh\alpha x}{\cosh\alpha x}\\
\\
\displaystyle\frac{\sqrt{\alpha}}{2}\frac{1}{\cosh\alpha x}
\end{array}
\right).
\label{psi_s2_e3}
\end{equation}

Substituting (\ref{psi_s2_e3}) and $\phi(x)=\tanh\alpha x$ into (\ref{eqmo2}), we get the desirable self-consistency equation for the slope $\alpha$:
\begin{equation}
2\alpha^2-2=-\sqrt{3}\alpha^2 \quad\Longrightarrow\quad \alpha^2=\frac{2}{2+\sqrt{3}},
\label{self_con}
\end{equation}
hence, $\alpha=\sqrt{2/(2+\sqrt{3})}\approx 0.732$.
With this value for $\alpha$ the distorted kink $\tilde{\phi}_K(x)=\tanh\alpha x$ bounds the excited fermion with the wave function given by (\ref{psi_s2_e3}).
This is an example of a self-consistent solution of the problem ''kink + excited non-zero frequency fermion''.

\section{Conclusion}

\setcounter{equation}{0}

The procedure we used here is similar to the Hartree-Fock method widely used in the atomic and nuclear physics. The main our result is the self-consistency condition (\ref{self_con}),
which is a purely algebraic one in our case. The slope of the distorted kink at the origin in the presence of valence fermion changes from $\alpha=1$ to $\alpha\approx 0.732$.

For the particular value of $s$, $s=2$, (and/or $g=2\alpha$) the distorted kink can capture a couple of fermions, one being in the ground state (zero mode) and
the second one in the first excited state with the wave function given by eq.~(\ref{psi_s2_e3}).
If a couple of fermions occupies both excited states with $\varepsilon^\prime=\pm\sqrt{3}$, the solution of the self-consistency condition for $\alpha$ is
$\tilde{\alpha}=1/\sqrt{1+\sqrt{3}}\approx 0.605$.

Note that the solution of the problem with negative $\alpha=-\sqrt{2/(2+\sqrt{3})}$ also exists. It corresponds to the case of a distorted antikink coupled to an antifermion.

For arbitrary $s$ the wave function of fermion for the first excited state in the field of kink $\tilde\phi_K(x)=\tanh(\alpha x)$ reads:
\begin{equation}
\psi_{\varepsilon^\prime}(x)=
\left(
\begin{array}{l}
u_{\varepsilon^\prime}(x)\\
v_{\varepsilon^\prime}(x)
\end{array}
\right)=
\sqrt{\frac{\alpha\:\Gamma(s-1/2)}{2\sqrt{\pi}\:\Gamma(s-1)}}
\left(
\begin{array}{c}
\displaystyle\sqrt{2s-1}\ \frac{\tanh\alpha x}{\cosh^{s-1}\alpha x}\\
\\
\displaystyle\frac{1}{\cosh^{s-1}\alpha x}
\end{array}
\right),
\label{psi_arb_s}
\end{equation}
where $\Gamma(z)$ is the gamma-function. Substituting (\ref{psi_arb_s}) into eq. (\ref{eqmo2}) we obtain the following constraint:
\begin{equation}
2\alpha^2-2=-\frac{\alpha^2 s\:\Gamma(s-1/2)\sqrt{2s-1}}{\sqrt{\pi}\:\Gamma(s-1)}\frac{1}{\cosh^{2s-4}\alpha x}.
\label{eq_arb_s}
\end{equation}
In the limit $s\to 1$ (\ref{eq_arb_s}) reproduces $\alpha\equiv 1$, $\phi_K=\tanh x$ what means that we have only one bound state in the system, i.e. the first
excited state for fermion belongs to the continuum. As it follows from eq. (\ref{eq_arb_s}),
for any $s\ne 1$ and $s\ne 2$ the solution for scalar field in the form of distorted kink with constant $\alpha$ does not exist.
However for the case $|s-2|\ll 1$ the solution for kink's profile $\tilde{\phi}_K=\tanh\alpha x$ with $\alpha=1/\sqrt{1+\sqrt{3}/2}$
looks quite reasonable at small distances $x$
satisfying the condition
\begin{equation}
|s-2|\cdot x^2 \alpha^2\ll 1.
\end{equation}

Speaking of the practical applications of the results discussed above, we shall first of all mention the problems of cosmology. In this connection we refer to the studies of Refs.~\cite{Zel}.

\section {Acknowledgments}
The authors are thankful to Dr. V.~A.~Lensky for careful reading of the manuscript and for many useful remarks.
This work was partially supported by the Russian State Atomic Energy Corporation ''Rosatom''
and by grant NSh-4568.2008.2.

\end{document}